\documentclass[twocolumn,reprint,epsf,graphics,psfig,superscriptaddress]{revtex4}

\usepackage{amsfonts}
\usepackage{graphicx}
\usepackage{amsmath}
\usepackage{amssymb}
\usepackage{latexsym}
\usepackage{psfrag}
\usepackage{dcolumn}
\usepackage{datetime}
\usepackage{multirow}
\usepackage{subfigure}
\usepackage{amssymb,amsmath,rotate,color}

\begin{document}

\title{Gas adsorption effects on electronic and magnetic properties of triangular graphene antidot lattices}
\author{Zahra Talebi Esfahani}
\affiliation{Department of Physics, Payame Noor University, P.O.Box 19395-3697 Tehran, Iran}
\author{Alireza Saffarzadeh}
\affiliation{Department of Physics, Payame Noor University, P.O.Box 19395-3697 Tehran, Iran} 
\affiliation{Department of Physics, Simon Fraser University, Burnaby, British Columbia, Canada V5A 1S6}
\author{Ahmad Akhound}\email{aakhound@yahoo.com}
\affiliation{Department of Physics, Payame Noor University, P.O.Box 19395-3697 Tehran, Iran}
\author{Amir Abbas Sabouri Dodaran}
\affiliation{Department of Physics, Payame Noor University, P.O.Box 19395-3697 Tehran, Iran}
\date{\today}

\begin{abstract}
The adsorption effects of small molecules (H$_{2}$O, CO, NH$_{3}$, NO$_{2}$) and large molecules (Tetracyanoquinodimethane (TCNQ) and Tetrafluoro-tetracyanoquinodimethane (F4TCNQ)) on electronic and magnetic properties of two triangular graphene antidot lattices (GALs), $[10,3,6]_{RTA}$ and $[10, 5]_{ETA}$, are investigated by means of first-principles calculations. We find that CO, NO$_{2}$, TCNQ, and F4TCNQ molecules are chemisorbed by both antidots, whereas NH$_{3}$ is physisorbed (chemisorbed) by $[10, 5]_{ETA}$ ($[10,3,6]_{RTA}$) structure. H$_{2}$O, CO, NH$_{3}$ molecules reveal no significant effect on electronic and magnetic properties of these antidot structures. The adsorbed NO$_{2}$ molecules affect the energy gap of GALs by changing their electronic structure from semiconducting to half-metal nature. This suggests that both GALs can act as efficient  NO$_{2}$ sensors. The adsorption of TCNQ and F4TCNQ molecules on GALs induces flat bands in the vicinity of the Fermi energy and also turn the electronic structure of antidot lattices to half-metallicity. Among the small and large molecules, NO$_{2}$ molecules induce the most total magnetic moment, paving the way to make magnetic GAL-based devices.

\end{abstract}

\maketitle

\section{Introduction}
Graphene, as a single atomic layer of graphite, has been the subject of a large amount of theoretical and experimental studies during the last decades, because of its rich and fascinating physical properties \cite{nov1,cas2,nov3,bol4,nai5} which can lead to many novel applications from more efficient solar cells to medicinal technologies. For instance, due to the Dirac-like spectrum of charge carriers in its gapless band structure, the electronic properties of graphene have attracted a great deal of research interest. Nevertheless, the zero band gap nature of graphene limits its practical applications in optoelectronics and photonics. To open an energy gap at Dirac point one can cut graphene into nanoribbons, create defects or add gas atoms \cite{lee7,wan8}. One way to create defects is to remove some atoms and make a periodic array of holes, known as graphene antidot lattice (GAL) \cite{fur9,van10,ped11,ouy12,pet13,ped14,pow15}. 

The electric and magnetic properties of some GALs have been extensively studied in recent years \cite{ped11,ouy12,pet13,ped14,pow15,fur16,van17,our18,sch23}. 
Moreover, the investigation of sensing properties of GALs in the presence of small and large gas molecules may promote our fundamental understanding of the underlying physics and their possible technological applications. In this regard, Brun et al. \cite{sor34} studied the adsorption of boron and nitrogen doping on the hexagonal antidots and that how the size of antidot supercell and the number of passivated atoms affect the GAL properties.
  
The small molecules, such as CO, H$_{2}$O, NH$_{3}$ and NO$_{2}$  are common adsorbates on graphene substrate \cite{Schedin2007,lee24,you25,lin26}. It has been shown experimentally \cite{Schedin2007} that to make highly sensitive graphene-based sensors, one can increase the graphene charge carrier concentration by adsorption of these small gas molecules. Moreover, using the density functional theory (DFT), it has been reported that H$_{2}$O and NO$_{2}$ molecules on graphene act as acceptors, while NH$_{3}$ and CO molecules exhibit electron donors \cite{lee24}. Among these adsorbates, the paramagnetic molecule NO$_{2}$ is a strong dopants \cite{lee24}. The charge transfer between the gas molecules and graphene surface depends strongly on the orientation of the adsorbate with respect to the surface but it is almost independent of the adsorption site. On the other hand, F4TCNQ and TCNQ are large molecules with high electron affinity \cite{ger,men,hsi,bal27,pin28,xia29}. TCNQ is an organic compound with chemical formula C$_{12}$N$_{4}$H$_{4}$ acting as an electron acceptor. This molecule which forms charge-transfer salts has received a great deal of attention due to its high electrical conductivity at room temperature \cite{Ferraris1973}. F4TCNQ molecule with chemical formula C$_{12}$N$_{4}$F$_{4}$  has the same TCNQ structure and only the four hydrogen atoms are replaced by fluorine (see Fig. 1(c) and (d)). Since fluorine has higher electronegativity than hydrogen, F4TCNQ is more electronegative than TCNQ. Both of these adsorbates are strong electron acceptors. As a result, the adsorption of these large molecules on graphene surface modifies strongly the electric and magnetic properties of the surface \cite{ger,men,hsi,bal27,pin28,xia29}. Nevertheless, the interplay between these molecules and GALs has, to our knowledge, not been reported previously.
 
\begin{figure}
\centerline{\includegraphics[width=0.95\linewidth]{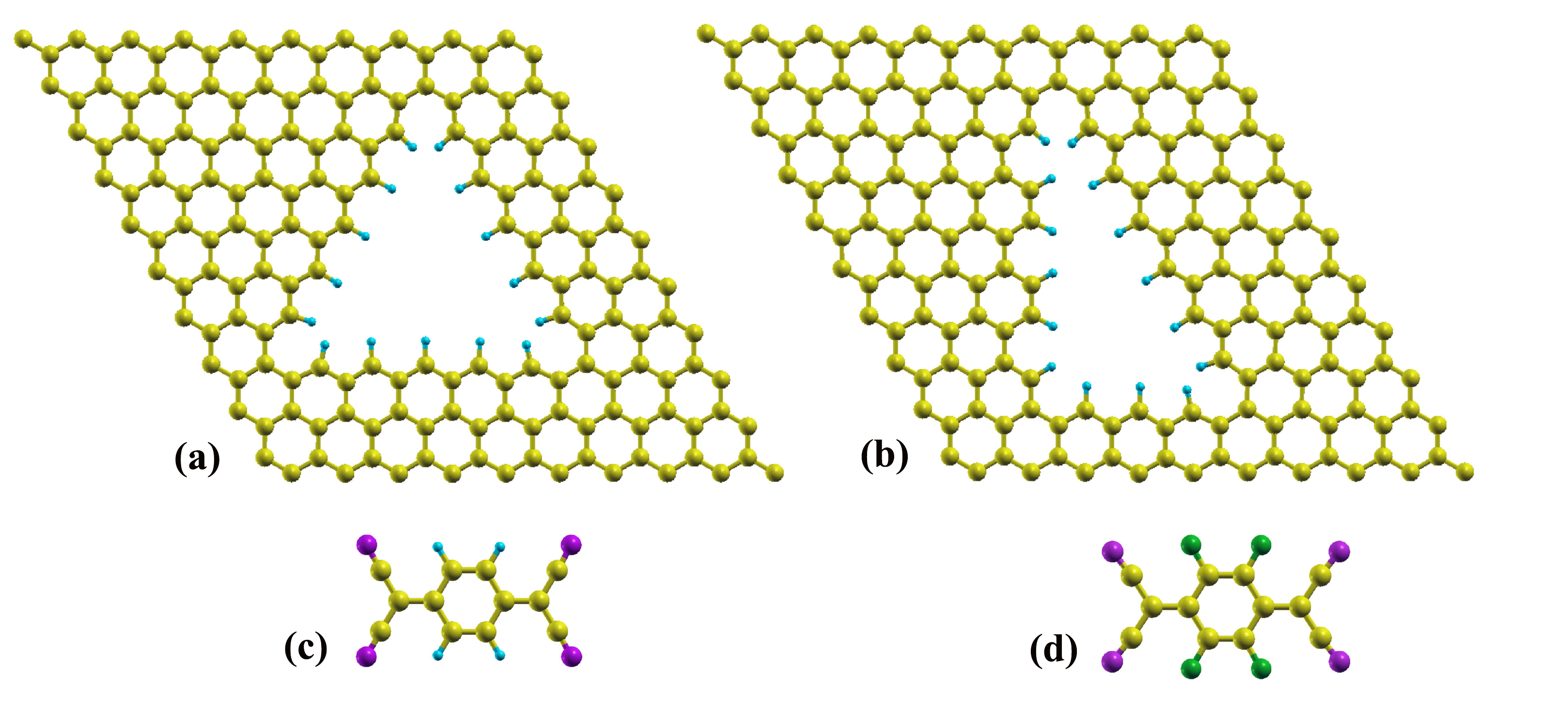}}
\caption{Optimized geometry of (a) $[10,5]_{ETA}$, (b) $[10, 3,6]_{RTA}$ supercells and (c) TCNQ and (d) F4TCNQ molecules. The yellow, blue, purple, and green balls represent carbon, hydrogen, nitrogen, and fluorine atoms, respectively.}
\label{image1}
\end{figure}

In this paper, we explore the adsorption of small and large molecules on the surface of two different shapes of triangular GALs, namely, $[10,3,6]_{RTA}$ and $[10,5]_{ETA}$. The type of gas adsorption, semiconducting properties and magnetic moments of the antidot lattices in the presence of H$_{2}$O, CO, NH$_{3}$, NO$_{2}$, TCNQ, and F4TCNQ molecules are studied by means of density functional theory (DFT) calculations. To do this we present our computational method in section II. The numerical results of band structures and induced magnetic moments for the adsorption of gas molecules on the antidot lattices are given in section III. Finally, we present a brief conclusion in section IV. 
\begin{figure}
\centerline{\includegraphics[width=0.95\linewidth]{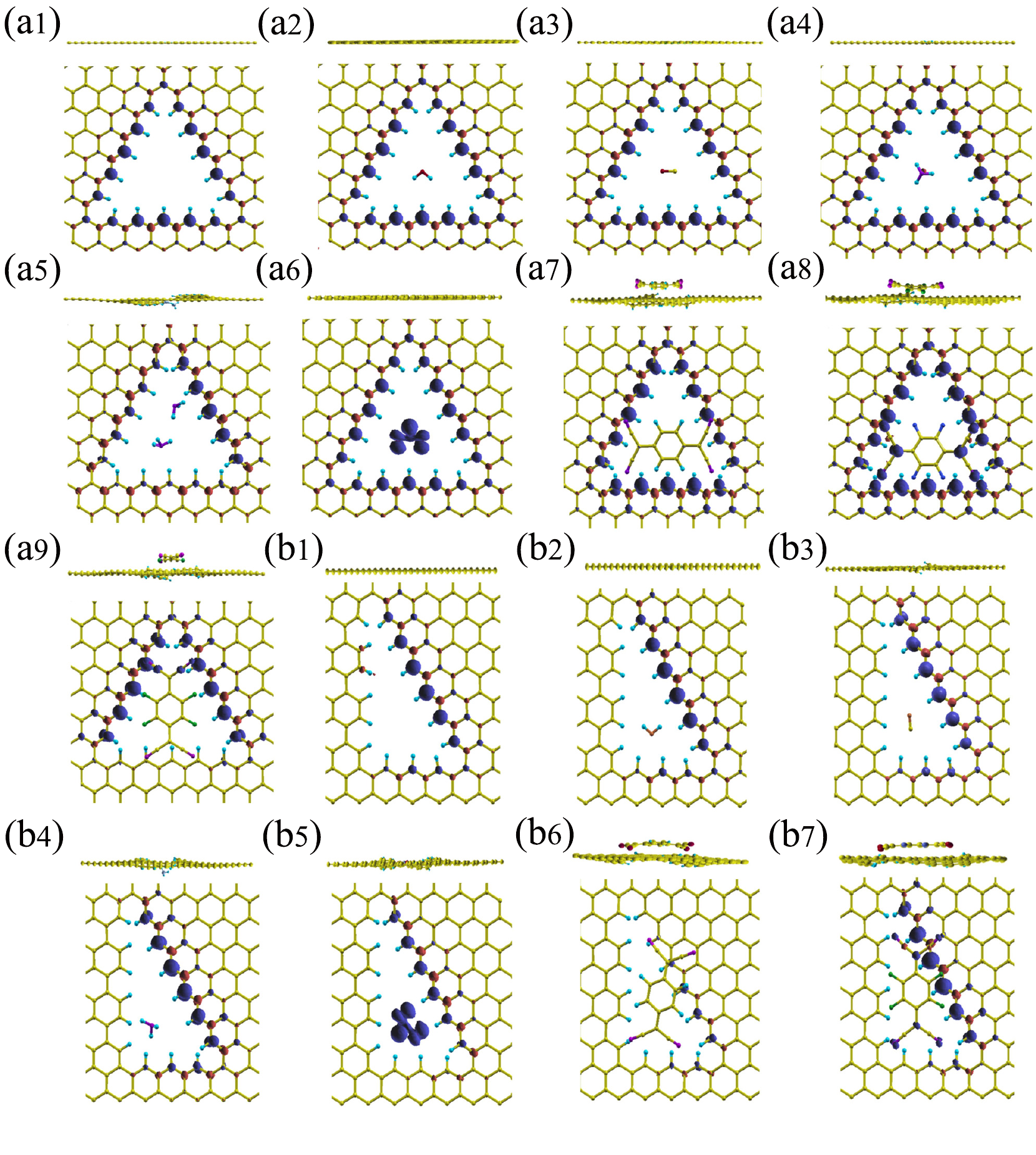}}
\caption{ The front view and side view (on top) of GALs in the presence of small and large molecules. (a1) pristine $[10,5]_{ETA}$, after adsorbing (a2) H$_{2}$O, (a3) CO, (a4) NH$_{3}$, (a5) 2NH$_{3}$, (a6) NO$_{2}$, (a7) TCNQ, (a8) F4TCNQ, (a9) F4TCNQ$_{90^{\circ}}$ molecules, and (b1) pristine $[10,3,6]_{RTA}$, after adsorbing (b2) H$_{2}$O, (b3) CO, (b4) NH$_{3}$, (b5) NO$_{2}$, (b6) TCNQ, and (b7) F4TCNQ molecules. The blue and red regions around each atom represent the charge densities corresponding to spin-up and spin-down electrons, respectively, with isosurface value 0.003 e/{\AA}$^{3}$.}
\label{image2}
\end{figure}

\section{Computational methods}
To investigate the electronic and magnetic properties of triangular GALs in the presence of gas molecules, we use two notations, $[L,D]_{ETA}$ for equilateral triangular antidot ( ETA) and $[L,Z,A]_{RTA}$ for right triangular antidot (RTA) patterns, created in the surface of graphene layers. 
We choose triangular antidot lattices, because the edge states of these structures are highly sensitive to the spatial arrangement of the atoms \cite{pow15,faragh2014} and induce magnetic moments on the zigzag-shaped edges in graphene nanoribbons and nanorings \cite{faragh2013,Fujita1996,Grujic2013}. Therefore, by examining the triangular antidots with pure zigzag edges (i.e., ETAs) and also the antidots with mixed armchair and zigzag edges (i.e., RTAs) in graphene lattices and comparing their spin-dependent electronic structures, we may provide a guideline for the future magneto-optical experiments in graphene-based nanodevices. In this regard, the triangular antidots in graphene have recently been fabricated \cite{Autin2016}, and experiments suggest the possibility of nanostructured zigzag-edged devices \cite{Bai2010,Shi2011,Oberhuber2013,Stehle2017}. Here $L$ represents the number of carbon atoms along the supercell edges, $D$ denotes the number of passivated carbon atoms along the ETA sides, and $Z$ and $A$ express the number of passivated carbon atoms along the two perpendicular sides of RTA with zigzag and armchair edges, respectively \cite{our18}. In this study, we consider superlattices with $L=10$, $D=5$, and $L=10$, $Z=3$, $A=6$, as shown in Fig. 1(a) and (b). 

The DFT calculations are performed using the SIESTA code \cite{sol35,ord36}. The exchange-correlation energy is calculated within the generalized gradient approximation (GGA) \cite{per37}. The Perdew-Burke-Ernzerhof (PBE) exchange-correlation functional, norm-conserving Troullier-Martins pseudo potentials, and a double-polarized basis (DZP) are used for this calculation \cite{art38}. 
Here, we should point out that although the PBE functional underestimates the electronic band gap \cite{Cohen2008,Mori2008}, it accurately describes the ground state properties. We use a $10 \times10$ ($L = 10$) supercell geometry with basis vectors $a$(1, 0, 0) and $a$($\frac{1}{2},-\frac{\sqrt{3}}{2},0$) in the $xy$ plane where $a = 24.63${\AA}. In addition, a vacuum space of size $a$ is applied along the $z$-axis to avoid the interactions between periodic images in the lattice. All structures are fully relaxed until the forces on each atom become smaller than 0.06 eV {\AA}$^{-1}$. The cut-off energy is set to 300 Ry and the Brillouin zone sampling is performed by the Monkhorst-pack mesh of $\bf k$-points. A mesh of $(10 \times 10 \times 2)$ has been adopted for discretization of ${\bf k}$-points. The spin-polarized calculations are performed to obtain the total magnetic moment. Also, the Mulliken population analysis is used in the calculation of local magnetic moment of each carbon atom. 

The adsorption energy, $E_{ads}$, of a single gas molecule is calculated using the following formula \cite{dam39}
\begin{equation}\label{eq1}
E_{ads} = E_{\mathrm{GAL+molecule}}- (E_{\mathrm{GAL}}- E_{\mathrm{molecule}})\ ,
\end{equation}
where $E_{\mathrm{GAL+molecule}}$ is the optimized energy of the GAL in the presence of adsorbed molecule, $E_{\mathrm{GAL}}$ and $E_{\mathrm{molecule}}$ are the optimized energies of the pristine GAL and the isolated molecule, respectively \cite{dam39,tay40}. It is worth mentioning that our calculations are limited to zero temperature. However, we should take the entropy corrections into account, when the gas adsorption is investigated at finite temperature.

Note that the SIESTA code calculates the Mulliken charge population for every single atom in the gas molecule before and after the adsorption. Therefore, the charge population determines whether the adsorbed molecule turns the system as a $p$-type (acceptor) or $n$-type (donor) semiconductor \cite{oli41}. The adsorption can be classified into physisorption (week van der Waals interaction) or chemisorption (covalent bonding). 
In general, the chemisorption is defined when the absolute value of adsorption energy is greater than 0.2 eV and the distance between the adsorbate and adsorbent surface is not large, otherwise the process is called physisorption \cite{XChen2017,tin30}. Here, we use the same definition for both small and large adsorbates. 

\section{Results}
We start with the small adsorbates H$_{2}$O, CO, NH$_{3}$ and NO$_{2}$ on $[10,3,6]_{RTA}$ and $[10,5]_{ETA}$ lattices. Each of these molecules are placed in the middle of the respective triangular hole created in the $10 \times10$ superlattices. Then the structures are optimized by means of the procedure explained above. From the  optimized structures we found that the $[10,5]_{ETA}$ lattice does not exhibit any structural distortion in the presence of gas molecules.  In other words, the small gas molecules are not able to bend or distort the pristine ETA lattices, and as a result, the corresponding lattice remains flat. The optimized superlattices before and after introduction of a small gas adsorbate are shown in Figs. 2(a1)-(a4) and 2(a6). The same procedure happens for  the $[10,3,6]_{RTA}$ lattice before and after introduction of H$_{2}$O molecules as shown in Fig. 2(b1) and (b2), while the structure exhibits a small distortion in the presence of CO molecule, as depicted in Fig. 2(b3).  
On the other hand, the optimized structures with other adsorbed molecules show a considerable out of plane distortion as can be seen in Fig. 2(a7)-(a9) and also Fig. 2(b4)-(b7). Such distortions in RTA lattices may happen due to the lack of structural symmetry, and also, as a result of the introduction of large molecules in both RTA and ETA lattices. The large gas molecules are placed over the hole at the distance of 3{\AA} from the plane of GAL. The ETA lattices possess zigzag edges only, while the RTA structures have the mixed armchair and zigzag edges. Therefore, the adsorbates in $[10,3,6]_{RTA}$ lattice move closer to the atoms of the hole edges compared to the $[10,5]_{ETA}$ lattice. Note that in Fig. 2(a5), we have also shown the optimized structure of $[10,5]_{ETA}$ lattice with two adsorbed NH$_{3}$ molecules, causing a considerable out of plane distortion.

In Table \ref{Table1}, the values of magnetic moment, adsorption energy, energy gaps, and the type of adsorbate are listed for both RTA and ETA lattices in the presence of small and large gas molecules. The adsorption energy of all structures is negative indicating that all optimized structures are stable. We discuss the effect of each adsorbate over GAL in details below.  

\begin{table*}[ht] 
\caption {Total magnetic moment $M_t$, adsorption energy $E_{ads}$, energy gap ($E^{\sigma}_{gap}$) for spin-$\sigma$ electrons, magnetic moment of each adsorbate $M_{a}$, and adsorbate type for different lattice structures. }
\centering
\begin{tabular}{|c|c|c|c|c|c|c|}
\hline
Structure & $~M_{t}(\mu_B)~$ & $~~~E_{ads}$(eV)~~ & $E_{gap}^{\uparrow}$(eV) & $E_{gap}^{\downarrow}$(eV) & 
$M_{a}(\mu_{B})$ & Adsorbate type\\
\hline
$[10,3,6]_{RTA}$ & 2.954 & & 0.517 & 0.550&&\\ 
\hline
$[10,3,6]_{RTA}$+H$_{2}$O & 2.943 & -0.523 & 0.512 & 0.543 & 0 & Donor \\
\hline
$[10,3,6]_{RTA}$ + CO & 2.965 & -1.730 & 0.532 & 0.555 & 0 & Donor\\
\hline
$[10,3,6]_{RTA}$ + NH$_{3}$ & 2.974 & -2.745 & 0.529 & 0.542 & 0 & Donor \\
\hline
$[10,3,6]_{RTA}$ + NO$_{2}$ & 2.723 & -3.855 & & 0.168 &1 & Acceptor \\
\hline
$[10,3,6]_{RTA}$ + TCNQ &1.012 & -5.149 & & 0.234 & 0 & Acceptor \\
\hline
$[10,3,6]_{RTA}$ + F4TCNQ & 3.003 & -5.313 & & 0.485 & 0 & Acceptor \\
\hline
$[10,5]_{ETA}$ & 4.984 & & 0.793 & 0.802 & &\\
\hline
$[10,5]_{ETA}$ + H$_{2}$O & 4.985 & -0.151 & 0.790 & 0.798 & 0 & \\
\hline
$[10,5]_{ETA}$ + CO & 4.986 & -0.428 & 0.794 & 0.815 & 0 & Donor \\
\hline
$[10,5]_{ETA}$ + NH$_{3}$ & 4.985 & -0.101 & 0.789 & 0.799 & 0 & \\
\hline
$[10,5]_{ETA}$ + 2NH$_{3}$ & 4.993 & -4.880 &0.804 & 0.823 & 0 & Donor \\
\hline
$[10,5]_{ETA}$ + NO$_{2}$ & 5.291 & -0.931 & & 0.572 & 1 & Acceptor \\
\hline
$[10,5]_{ETA}$ + TCNQ & 3.496 & -5.105 & & 0.245 & 0 & Acceptor \\
\hline
$[10,5]_{ETA}$ + F4TCNQ & 4.985 & -5.449 & & 0.680 & 0 & Acceptor \\
\hline
$[10,5]_{ETA}$ + F4TCNQ$_{90^{\circ}}$ & 3.040& -5.614 & & 0.281 & 0 & Acceptor\\
\hline 
\end{tabular}
\label{Table1} 
\end{table*}

\subsection{H$_{2}$O on $[10,3,6]_{RTA}$ and $[10,5]_{ETA}$ lattices }

The optimized structure of $[10,3,6]_{RTA}$ supercell in the presence of a single water molecule revealed that the molecule is located inside the hole at the distance of 1.97{\AA} from the edge atoms and that the calculated adsorption energy is -0.523 eV. On the contrary,  the adsorption energy of a single H$_{2}$O  on $[10,5]_{ETA}$ structure is -0.151 eV and the molecule is located at the distance of 2.72 {\AA} from the edge atoms (see Table 1). This shows that the binding interaction between water molecule and $[10,3,6]_{RTA}$ structure is much stronger than that between H$_{2}$O and $[10,5]_{ETA}$ lattice. Using the Mulliken charge analysis, one can conclude that H$_{2}$O molecule in $[10,3,6]_{RTA}$ lattice acts as a donor, while the amount of charge transfer  from $[10,5]_{ETA}$ lattice is very small. As shown in Figs. 3(b) and 4(b), the band structure of both lattices is not affected by adsorption of water molecules due to the weak bonding between H$_{2}$O and carbon atoms of the GAL, in agreement with the intrinsic hydrophobicity of graphene and graphitic surfaces \cite{Hong2016,LiuLi2017}.  The water molecule adsorption on the antidot lattices does not change the total magnetic moment of the respective pristine lattices, as given in Table. 1. Moreover, the distribution of charge densities remains unchanged if one compares Fig. 2(a2) with (a1) and Fig. 2(b2) with (b1). 

It is worth mentioning that H$_{2}$O molecule over graphene surface acting as an acceptor results an adsorption energy of -0.047 eV with the bond length of 3.5{\AA} \cite{lee24}. The water adsorption on antimonene (InS) which also behaves as an acceptor (acceptor) gives the adsorption energy of -0.20 eV (-0.17 eV) with the bond length of 2.88 {\AA} (2.37 {\AA}) \cite{oli41,ant-and}. On the contrary, the binding energy of H$_{2}$O over phosphorene is -0.14 eV with the bond length of 2.71{\AA} and for this case the molecule acts as a donor \cite{p-cai}.

\subsection{CO on $[10,3,6]_{RTA}$ and $[10,5]_{ETA}$ lattices }

The adsorption of carbon monoxide molecule, CO, on the GALs results the adsorption energies of -1.730 eV in $[10,3,6]_{RTA}$ and -0.428 eV in $[10,5]_{ETA}$ lattices, suggesting that CO molecule is chemisorbed by both lattices (see Table I). Also, the molecule is located at the distance of 2.25{\AA} and 2.67{\AA} from the antidot edges of $[10,3,6]_{RTA}$ and $[10,5]_{ETA}$ lattices, respectively. Figures 3(c) and 4(c) show that the adsorption of CO molecule has a negligible effect on the gap and energy bands of the GALs.  

Our Mulliken charge analysis showed that CO molecule acts as a donor in both GAL structures, that is the same as graphene and phosphorene \cite{lee24,p-cai}. Nevertheless, the binding energy of the molecule at the distance of 3.74{\AA} (3.06{\AA}) above pristine graphene (phosphorene) is -0.014 eV (-0.31 eV) \cite{lee24,p-cai}. For comparison, its adsorption energy above InS (antimonene) at the distance of 3.08{\AA} (3.72{\AA}) is -0.13 eV (-0.12 eV) \cite{in-cai,ant-and}. However, the charge transfer between the molecule and both InS and antimonene surfaces is very small \cite{in-cai,ant-and}. The distribution of charge densities is not affected by the adsorption of CO molecule on both GAL structures. This can be confirmed by comparing the blue and red regions of Fig. 2(a3) with 2(a1), and that of 2(b3) with 2(b1).
\begin{figure}
\centerline{\includegraphics[width=0.98\linewidth]{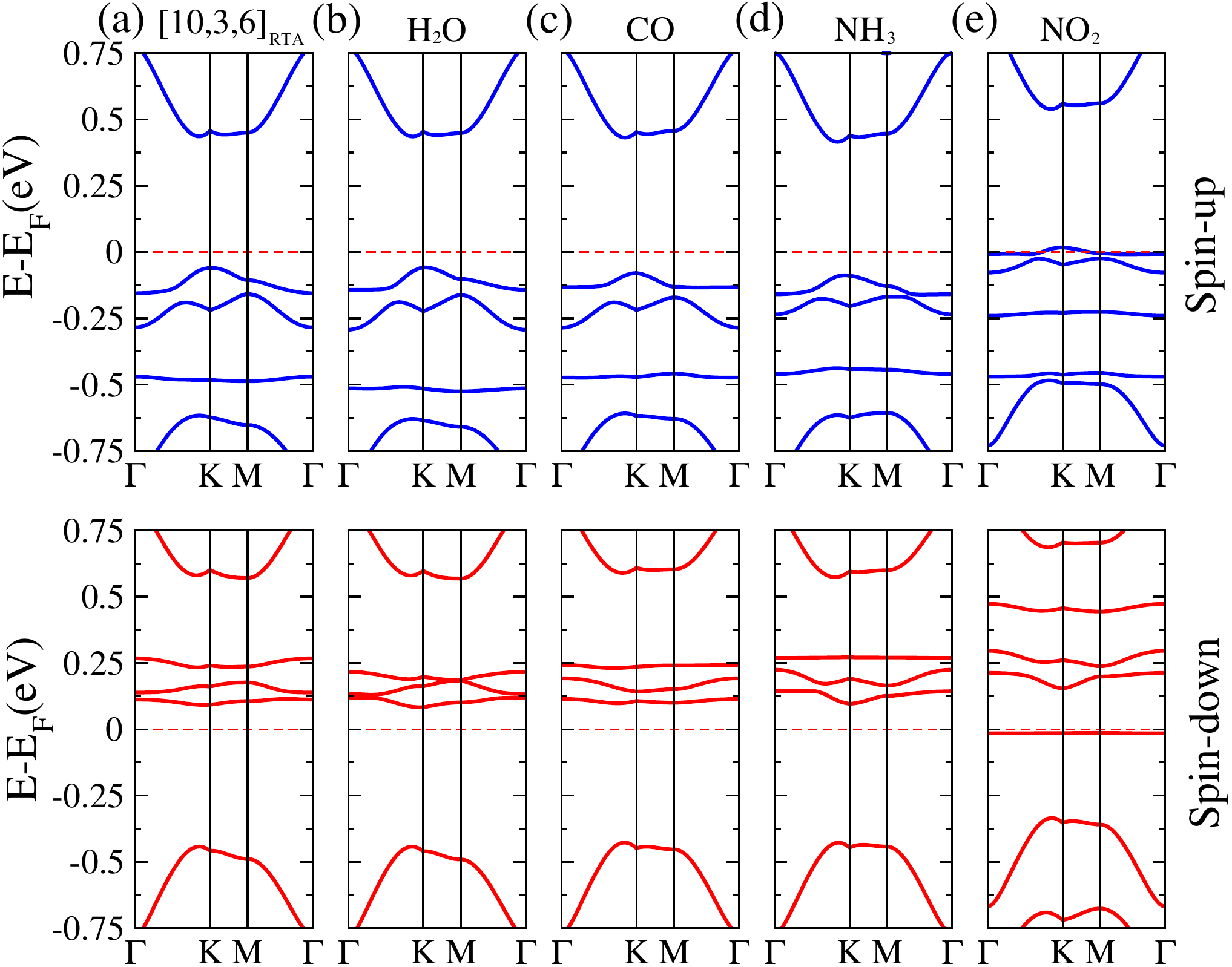}}
\caption{ Spin-dependent band-structure of (a) pristine $[10,3,6]_{RTA}$ lattice and the same structure after adsorption of the small gas molecules (b) H$_{2}$O, (c) CO, (d) NH$_{3}$, and (e) NO$_{2}$.} 
\label{image3}
\end{figure}
\begin{figure}
\centerline{\includegraphics[width=0.98\linewidth]{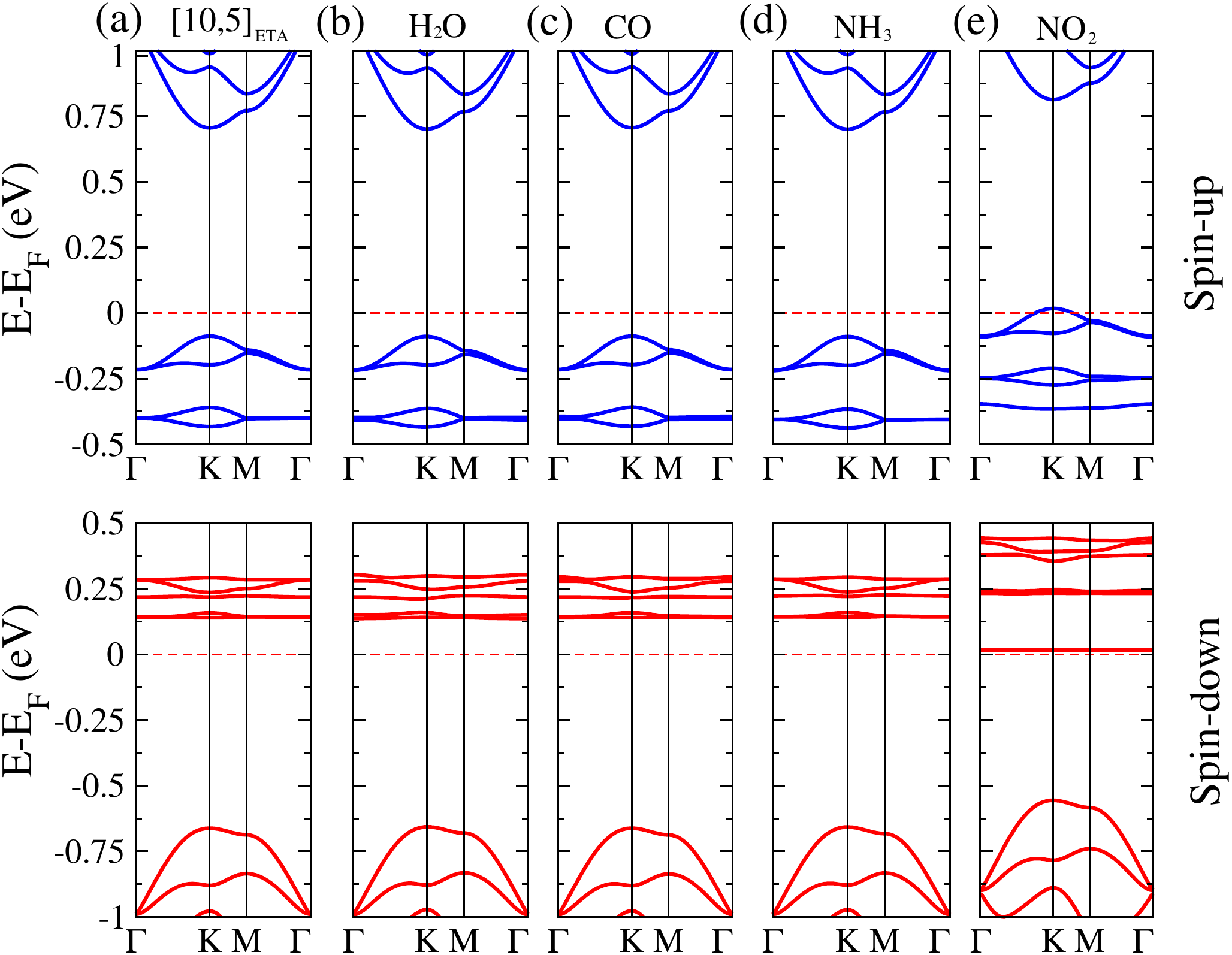}}
\caption{Spin-dependent band-structure of (a) pristine $[10,5]_{ETA}$ lattice and the same structure after adsorption of the small gas molecules (b) H$_{2}$O, (c) CO, (d) NH$_{3}$, and (e) NO$_{2}$.}
\label{image4}
\end{figure}

\subsection{NH$_{3}$ on $[10,3,6]_{RTA}$ and $[10,5]_{ETA}$ lattices }

Our ab initio calculations showed that the adsorption energy and the position of NH$_{3}$ molecule on $[10,3,6]_{RTA}$ are -2.75 eV and 2.192{\AA}, while these values are -0.1 eV and 2.05{\AA} for $[10,5]_{ETA}$ lattice (see Table I). This means that NH$_{3}$ molecule is chemisorbed by $[10,3,6]_{RTA}$ structure, while it is physisorbed by $[10,5]_{ETA}$ lattice. Again, the adsorption of NH$_{3}$ molecule does not change the energy bands of the GALs, as can be seen in Figs. 3(d) and 4(d).

The Mulliken charge analysis showed that the amount of charge transfer between NH$_{3}$ molecule and $[10,5]_{ETA}$ is very small but it is considerable in the case of $[10,3,6]_{RTA}$ lattice where the molecule acts as a donor. This situation is the same as NH$_{3}$ adsorption over graphene and phosphorene \cite{lee24,p-cai}. It has been reported that the adsorption energy of NH$_{3}$ molecule at the distance of 3.81{\AA} (2.59{\AA}) above graphene (phosphorene) surface is -0.031 eV (-0.18 eV) and the molecule acts as a donor \cite{lee24,p-cai}. On the other hand, the adsorption energy of the molecule at 2.49{\AA}  (3.41{\AA}) above InS (antimonene) is -0.20 eV (-0.12 eV) and the molecule acts as an acceptor \cite{in-cai,ant-and}.

In the case of small adsorbates, we have also examined the adsorption effect of two NH$_{3}$ molecules on $[10,5]_{ETA}$ holes by optimizing the total structure (see Fig. 2(a5)). The Mulliken charge analysis showed that the tendency of the molecules for charge transfer becomes smaller, while the adsorption energy increases significantly. Note that the gap value and the energy bands (not shown here) are not considerably affected by the presence of two NH$_{3}$ molecules.

\subsection{NO$_{2}$ on $[10,3,6]_{RTA}$ and $[10,5]_{ETA}$ lattices }
Among the various small adsorbates that we consider here, NO$_{2}$ molecule exhibits a nonzero magnetic moment in both GALs (see table \ref{Table1}). The distributions of charge densities in $[10,3,6]_{RTA}$ and $[10,5]_{ETA}$ lattices and also in the adsorbed molecule are shown in Fig. 2(a6) and (b5). The total magnetic moment is maximum in $[10,5]_{ETA}$ lattice while the molecule causes a reduction in the total magnetic moment of $[10,3,6]_{RTA}$ lattice. The adsorption energy and the position of NO$_{2}$ molecule on $[10,3,6]_{RTA}$ lattice are -3.855 eV and 2.03{\AA}, while these values are -0.931 eV and 2.66{\AA}\, for $[10,5]_{ETA}$ lattice. The energy values predict that the molecule is chemisorbed by both GALs.

As shown in Figs. 3(d) and 4(d), the Fermi energy crosses the spin-up energy bands of both GALs, whereas it falls in the band gap of spin-down band-structure with the energy gap value $\sim 0.168$ eV for $[10,3,6]_{RTA}$ and $\sim 0.572$ eV for $[10,5]_{ETA}$. This suggests that adsorbing NO$_{2}$ molecules by the GALs may turn the semiconductor characteristics of these systems into a half-metallicity behavior. The energy gap of $[10,5]_{ETA}$ lattice is more than three times greater than that of $[10,3,6]_{RTA}$ lattice. Therefore, in the presence of adsorbed NO$_{2}$ molecule, it will be more easier for electrons to make a transition from valence band of the $[10,3,6]_{RTA}$ lattice compared to $[10,5]_{ETA}$ structure. 

In the context of half-metallicity, the energy gap and the electronic structure which are related to each other, play an important role. Therefore, to ensure that the half-metallicity behavior, seen in the GALs in the presence of NO$_{2}$ molecules and obtained by PBE (GGA) functional is reliable enough, we have also performed DFT+$U$ calculations for NO$_{2}$ adsorption over $[10,5]_{ETA}$ structure with several $U$ values for oxygen atoms. Our results showed that although some energy bands are shifted in energy as $U$ parameter changes, the energy gap value and also the energy bands around the Fermi energy are not affected by on-site Coulomb correction. Note that the GGA+$U$ method can be applied to any states which are too high (or shallow) in energy, too delocalized (due to the GGA-inherent self-interaction error), and hence, too metallic (the gap closes or becomes negative) \cite{Keating2012}.

In addition, the shift of Fermi level into the valence band of spin-up electrons indicates that NO$_{2}$ may act as a $p$-type impurity in GALs, similar to the case of graphene \cite{tay40}. However, the spin-down band-structures demonstrate a mid-gap state (flat band), induced by NO$_{2}$ molecule. The mid-gap state is an unoccupied state at 0.014 eV above the Fermi level in the $[10,5]_{ETA}$ lattice, whereas it is an occupied state at 0.014eV below the Fermi level in the $[10,3,6]_{RTA}$ lattice. Such a flat band has also been reported for the case of NO$_{2}$ molecule on monolayer of MoS$_{2}$ \cite{yue} and tetragonal GaN \cite{yon}. Therefore, the $[10,3,6]_{RTA}$ and $[10,5]_{ETA}$ lattices can act as viable NO$_{2}$ detectors. Note that the acceptor action of NO$_{2}$ molecules on GALs, has also been predicted for this adsorbed molecule on graphene ($d=3.38${\AA}\, and $E_{ads}=-0.067$ eV) \cite{lee24}, antimonene ($d= 2.44${\AA}, $E_{ads}=-0.81$ eV) \cite{ant-and}, phosphorene ($d=2.27${\AA} and $E_{ads}=-0.51$ eV) \cite{p-cai}, and InS ($d=2.71${\AA} and $E_{ads}=-0.24$ eV) \cite{in-cai}.

Note that to see to what extent the van der Waals (vdW) forces can affect the electronic structure of GALs in the presence of gas molecules, we have also included the vdW corrections in our band structure calculations of NO$_{2}$ molecule on the GALs. Our results showed that the energy bands (not given here) around the Fermi energy and the band gap values obtained from GGA calculations are not affected by the inclusion of vdW interactions, originated from quantum fluctuations of electric charge.

\begin{figure}
\centerline{\includegraphics[width=0.68\linewidth]{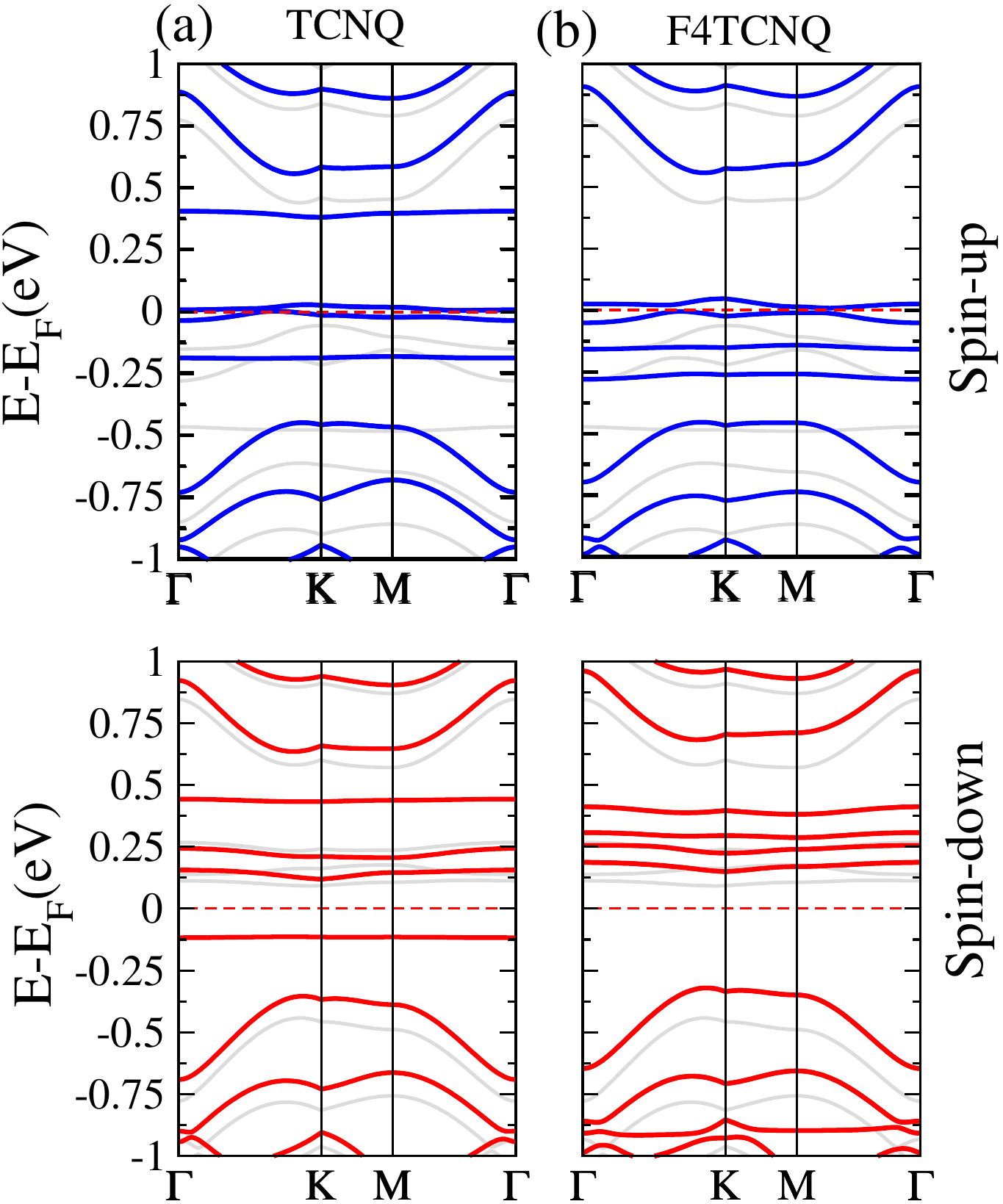}}
\caption{Spin-dependent band-structure of $[10,3,6]_{RTA}$ lattice before (grey curves) and after (blue and red curves) adsorbing the large molecules (a) TCNQ and (b) F4TCNQ. Blue (red) curves represent the energy bands of spin-up (spin-down) electrons.}
\label{image5}
\end{figure}
\begin{figure}

\centerline{\includegraphics[width=0.95\linewidth]{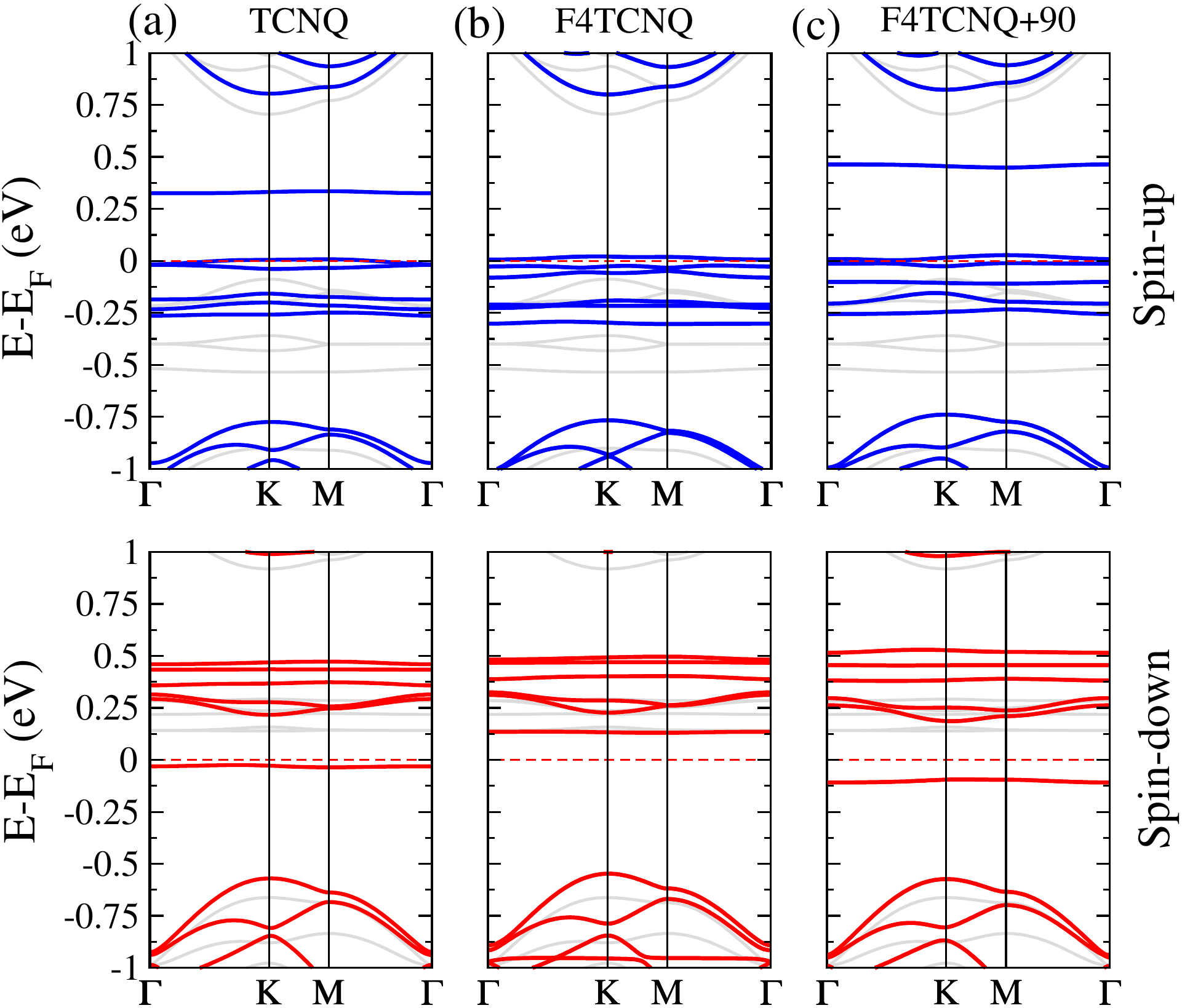}}
\caption{ Spin-dependent band-structure of $[10,5]_{ETA}$ lattice before (grey curves) and after (blue and red curves) adsorbing the large molecules (a) TCNQ, (b) F4TCNQ, and (c) F4TCNQ$_{+90^{\circ}}$. Blue (red) curves represent the energy bands of spin-up (spin-down) electrons.}
\label{image6}
\end{figure}

\subsection{TCNQ on $[10,3,6]_{RTA}$ and $[10,5]_{ETA}$ lattices }

The TCNQ molecule is a large organic adsorbate that is known as a strong electron acceptor. We obtained the energy adsorption of the molecule on the $[10,3,6]_{RTA}$ lattice $E_{ads}= -5.149$ eV and the shortest distance between the molecule and the GAL $d=2.17${\AA}, while the values are -5.105 eV and 2.07{\AA} for the $[10,5]_{ETA}$ structure, as given in Table I. We have depicted in Figs. 5(a) and 6(a) the band-structures of GALs after adsorption of TCNQ. As shown in the Table \ref{Table1}, the adsorbed molecule reduces the total magnetic moment of both structures, compared to the relative pristine lattices. The reduction in the magnetic moment of the $[10,5]_{ETA}$ structure is less than that of the $[10,3,6]_{RTA}$ lattice.
The adsorbed molecule induces several flat bands in the energy gap of both systems with a shift towards their valence band. This suggests the TCNQ molecule as a $p$-type impurity which can act as an acceptor in the GALs. The spin-up energy bands cross the Fermi level, while the spin-down band structures exhibit the energy gaps of 0.234 eV and 0.245 eV in $[10,3,6]_{RTA}$ and $[10,5]_{ETA}$ lattices, respectively. This behavior shows that the adsorption of TCNQ molecule by GALs may transform these semiconductors into half metals.

In this regard, Yang et al. \cite{yan} showed that the adsorption energy of TCNQ molecule on graphene varies from -1.19 eV to -1.53 eV by changing the calculation method and the orientation of adsorbate. Also, its adsorption energy over phosphorene and Al(100) reported about -0.97 eV \cite{yu} and -3.66 eV \cite{vin}, respectively. In these cases, the TCNQ molecule also acts as a strong acceptor. 

\subsection{F4TCNQ on $[10,3,6]_{RTA}$ and $[10,5]_{ETA}$ lattices} 

The F4TCNQ molecule is considered as an excellent $p$-type dopant on graphene surface \cite{wee} with adsorption energy ranging from -1.42 eV to -1.81 eV, depending on the calculation methods and the change in orientation of the molecule \cite{yan}. From optimized structures, we obtained $E_{ads}=-5.313$ eV and $d=2.13${\AA} in $[10,3,6]_{RTA}$, while $E_{ads}=-5.449$ eV and $d=1.99${\AA} in $[10,5]_{ETA}$ lattice. This means that F4TCNQ molecule can be considered as a stronger adsorbate in GALs, compared to TCNQ molecule. The band structures of both GALs in the presence of F4TCNQ are shown in Figs. 5(b) and 6(b). The Fermi energy of both lattices crosses the spin-up bands, while it falls in the gap of spin-down band structures with energy of 0.485 eV in $[10,3,6]_{RTA}$ lattice and 0.680 eV in $[10,5]_{ETA}$ lattice. Note that in both lattices the total magnetic moment is strong (see Table \ref{Table1} and also distribution of charge densities in Fig. 2(a8) and (b7)).

In the case of $[10,5]_{ETA}$ lattice, we have also considered the effect of 90$^{\circ}$ rotation of the molecule above the antidot, as shown in Fig. 2(a9). This rotation which is parallel to the GAL plane results $E_{ads}=-5.614$ eV and $d=2.36${\AA}. Comparing $E_{ads}$ values before and after 90$^{\circ}$ rotation of the molecule, we find that such a rotation can slightly increase the adsorption energy while it decreases the total magnetic moment considerably, as can be predicted from spin-dependent charge densities of Fig. 2(a9). The degeneracy of energy bands is broken slightly in the vicinity of the Fermi energy and a flat band is shifted above (below) the Fermi energy in the spin-up (spin-down) band structure, as depicted in Fig. 6(c). Nevertheless, the electronic structure of the system is not considerably affected by rotation of the molecule. The system is conductive for spin-up electrons, whereas it acts as a semiconductor with the energy gap of 0.281 eV for spin-down electrons. 

Therefore, the above features of GALs in the presence of large adsorbed molecules suggest a half-metallic nature with metallic behavior in one spin channel and insulating in the other. The energy bands of GALs are shifted towards higher energies by these molecules. In addition, the electronic and magnetic properties of these structures can be controlled by the type and orientation of the large molecules. The appropriate molecule can be selected by considering the required energy gap and the total magnetic moment. 

\section{conclusion}
In this paper we have studied the electric and magnetic properties of $[10,3,6]_{RTA}$ and $[10,5]_{ETA}$ lattices before and after adsorbing small and large gas molecules using DFT calculations, implemented in the SIESTA package with periodic boundary conditions. Our findings show that NH$_{3}$ is physisorbed (chemisorbed) by $[10, 5]_{ETA}$ ($[10,3,6]_{RTA}$) lattice, while CO, NO$_{2}$, TCNQ, and F4TCNQ molecules are chemisorbed by both lattices. The binding interaction between water molecule and $[10,3,6]_{RTA}$ lattice is stronger than that between H$_{2}$O and $[10,5]_{ETA}$ structure. Although the electronic and magnetic properties of these GALs are not considerably affected by H$_{2}$O, CO, NH$_{3}$ molecules, the  NO$_{2}$ adsorbate modulates the energy gap by transforming the electronic structure from semiconducting to half-metal nature. Also, the adsorption of TCNQ and F4TCNQ molecules on these lattices can turn the system into a half-metallicity as a result of shifting the energy bands around the Fermi energy. Our calculations suggest that the adsorption of NO$_{2}$ molecules with maximum induced magnetic moment, can be a promising candidate to design nanoscale spintronic devices and gas sensors, based on triangular graphene antidot lattices.

\end{document}